\documentclass[showpacs,preprintnumbers,amsmath,amssymb]{revtex4}

\usepackage{amsmath}

\usepackage{dcolumn}
\usepackage{bm}
\usepackage{rotating}
\usepackage{graphics}
\usepackage{color}

\begin{document}

\preprint{APS/123-QED}

\title{Highly Fluorinated Graphene}

\author{Chad E. Junkermeier}
\affiliation{NRC Research Associate at Naval Research Laboratory, Washington DC 20375}

\author{Stefan C. Badescu}
\affiliation{Air Force Research Laboratory, Wright-Patterson AFB, OH 45433}

\author{Thomas L. Reinecke\footnote{Corresponding author: reinecke@nrl.navy.mil}}
\affiliation{Naval Research Laboratory, Washington DC 20375}

\date{\today}



\begin{abstract}
We give results of density functional calculations for graphene with a wide variety fluorine adsorption patterns.  A systematic analysis of the adsorption energies, lattice constants, bulk moduli, band gaps, and magnetic moments is given. A range of different physical properties can occur for different adsorption geometries for each stoichiometry. For example, the systems are found to range from metallic to semiconducting with widely varying band gaps, and a number of interesting magnetic properties are found.  These results may help in guiding experiment and  materials development.
\end{abstract}

\maketitle

\section{Introduction}\label{sec:intro}

Since its experimental realization graphene has been under intense study because of its exceptional physical properties including high electrical conductivity, strength, and elasticity~\cite{Novoslev04666,Novoslev05197,Geim07183}.  However, pristine graphene is relatively inert chemically and is a zero band gap semiconductor; thus many of its properties need to be modified for use in applications~\cite{Wolf011488}.  Methods to modify the properties of graphene include substrate interactions~\cite{Giovannetti07073103,Sutter08406,Decker112291,Ong11075471,Xu112735}, electric and magnetic fields~\cite{Robinson083137,Yu093430,Fujita10043508}, strain~\cite{Fujita10043508,Low103551}, and chemical functionalization~\cite{Casolo09054704,Sofo11081411,Zhou09103108}.  Surface chemical functionalization is particularly attractive because graphene consists of only surface atoms.  A number of adsorbates have been studied, including hydrogen~\cite{Casolo09054704}, fluorine~\cite{Sofo11081411}, oxygen~\cite{Baraket103382,Robinson083137}, and primary amine (NH$_{2}$)~\cite{Baraket12233123,Junkermeierjp309419x}.  This work focuses on chemical functionalization with fluorine.

The fluorination of carbon-based systems has attracted considerable attention.  Much work has been directed at the fluorination of graphite, particularly in connection with lubrication~\cite{Kita793832}, and considerable experimental and theoretical work has been done on fluorination of buckyballs and carbon nanotubes~\cite{Dresselhaus932054}.  Fluorine adsorption is a relatively simple chemical functionalization that increases the reactivity of sp$^{2}$ bonded graphene and opens a wide range of modifications~\cite{Stine127957}, including tuning of band gaps and doping for use in electronics~\cite{Enoki03}.
Experiments have shown dramatic changes in the electronic and structural properties of graphene by increasing the degree of fluorination~\cite{Baraket10231501,Robinson103001,Nair102877,Zboril102885,Zalalutdinov124212,Robinson2012}.
Theoretical work on fluorinated graphene has focused mainly on fully covered systems~\cite{Medeiros10485701,Leenaerts10195436}, and on systems with low fluorine coverage~\cite{Boukhvalov11055708,Solenov5920}. Relatively few studies have considered high coverage fluorinated graphene systems~\cite{Ewels06216103,Boukhvalov08085413,Artyukhov105389,Ribas11143,Sahin11115432,liu1218193}.

Methods for controlling where and how adsorbates bind to graphene are being developed at both the micro and atomic scale.  Methods to control the micro scale patterns include: hot atom beams~\cite{Balog10315}, electrostatically biased scanning probes~\cite{Lee13s1227401303551}, laser-irradiation~\cite{Lee122374}, polymer masks~\cite{Lee115461}, and physical masks~\cite{Hernández1384}.   Theoretical research indicates it may be possible to tailor, at least to some degree, the atomic arrangement of adsorbates on the graphene basal plane by combinations of strain~\cite{Cole1111122677}, laser-induced desorption~\cite{Zhang12201409}, and changing the chemical potentials~\cite{Ao1014503,Jiang1219321,Solenov5920}.  Although the methods involved often do not produce well-ordered adsorbate structures, recent experimental evidence suggests that F can be adsorbed into a well-ordered structures that are not fully fluorinated~\cite{Schmucker2012,kashtiban2014atomically}.  These experimental methods and theoretical ideas open the possibility of designing fluorinated graphene systems with the patterning needed to obtain desired properties.

Here we consider the properties of systems with varying degrees of fluorination (stoichiometries) on graphene surfaces and with a number of different arrangements of F atoms (motifs) for each stoichiometry.  The overall stoichiometry of F on a fluorinated graphene surface is denoted by C$_{m}$F$_{n}$ where there are $n$ F atoms adsorbed onto $m$ C atoms.  In Section~\ref{sec:methods} we discuss the computational methods used here.  In Section~\ref{sec:results} the mechanical, electrical, elastic, and magnetic properties of fluorinated graphene of varying stoichiometries and motifs are presented.  A listing of the stoichiometries and motifs and their calculated properties is given in Table~\ref{table:configurations}.  We will frequently refer to this table.  Concluding remarks are found in Section~\ref{sec:Summary}.

\section{Computional Methods}\label{sec:methods}

We used the \textit{ab initio} plane wave code PWscf, which is part of Quantum Espresso~\cite{Giannozzi09395502}, to study supercells of fluorinated graphene with periodic boundary conditions.  The generalized gradient approximation of Perdew-Burke-Ernzerhof (PBE) was used for exchange-correlation effects.  Vanderbilt ultra-soft psuedopotentials~\cite{QE_HCF_UPF2010} with wave function energy cutoff of 30 Ry was employed.  We found that an 8-by-8 Monkhorst-Pack mesh sampling of the Brillouin zone was sufficiently dense for the total energy to be converged.  To speed up the self-consistent field theory (SCF) conversion during the structural relaxations the wave functions and potentials were extrapolated to second order from the preceding ionic steps.

\begin{figure}
\begin{center}
\includegraphics[clip,width=1.5 in, keepaspectratio]{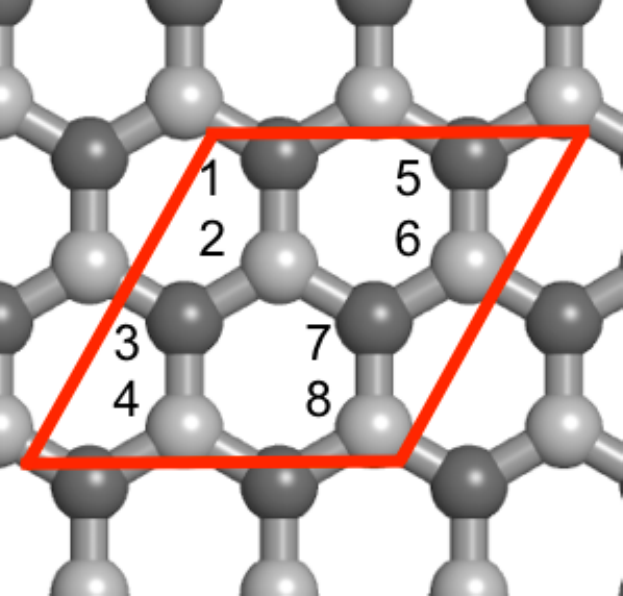}
\end{center}
\caption{Placement of carbon atoms in a two-by-two unit cell of graphene.  The numbers denote the carbon atoms in the unit cell that have fluorine atoms attached to them.  The dark (light) gray spheres represent the atoms in the A (B) graphene sublattice.  In one-sided adsorption, the attached atoms are directly over the carbon atoms coming out of the page.}
\label{fig:configurations}
\end{figure}

In this work we studied a large number of periodic systems using a 2-by-2-graphene cell as the basic unit.  We will refer to them by the Item numbers in the first column of Table~\ref{table:configurations}.  The second and third columns of Table~\ref{table:configurations} respectively give the stoichiometry and the coverage (percentage of C atoms with a F atom attached) of each system.  Figure~\ref{fig:configurations} labels the positions of the C atoms in the unit cell.  The starting configuration of each system had F atoms directly above, or below, the C atoms.  The motifs (arrangements) of the F atoms in each system are given in the columns labeled ``Above" and ``Below" in Table~\ref{table:configurations} where the numbers correspond to the C atom having a F adatom attached above or below the graphene plane~\cite{labelingexample}.  The calculated values of the lattice parameters were obtained by computing the total energy of the F-graphene system at a set of lattice constants and then fitting the resulting curve with the Birch-Murnaghan equation of state~\cite{Yoon2009,BMfitting}.  An inter-layer spacing (c lattice parameter) of 24.6 \AA\, was used in the direction perpendicular to the graphene plane~\cite{Artyukhov105389}.

For band structures and electronic properties we have taken into account the possibility of ferromagnetic and antiferromagnetic spin alignments.  In order to create ferromagnetic alignments all fluorine atoms were given an initial net moment in the same direction.  In the case of antiferromagnetic alignments, the fluorine atoms on the B sublattice were given initial moments in a direction opposite to those on the A sublattice of graphene~\cite{spinordering}.  The ordering with the lowest total energy is the ground state.  The adsorption energies, band gaps, intrinsic spin polarizations, and charge transfers in Table~\ref{table:configurations} and in the figures are the values found in the ground state configuration.

The calculated band structures of all C$_n$F phases studied here can be described by reduced sets of maximally localized Wannier functions, MLWF. These are localized, tight-binding-like Wannier orbitals obtained by band interpolation, and they are useful for understanding electronic transport~\cite{Calzolari04035108} and elasticity models~\cite{Straub8910325}. They describe the occupied bands and are a good approximation to the lowest conduction bands.  Wannier90 was used to produce MLWF from the Bloch-like wave functions used in Quantum Espresso~\cite{Mostofi08685,MLWF_uses}.  In performing the Wannierization, we specified the k-point positions of a 4x4 two-dimensional Monkhorst-Pack mesh sampling of the Brillouin zone, which was shifted to include the $\Gamma$-point.  The band disentanglement was performed over a k-point path from $M$, to $\Gamma$, to $K$.  In this way we can reproduce well the valence band of C$_{n}$F and the lowest branches in the conduction band.

\section{Results and Discussion}\label{sec:results}

\begin{figure}
\begin{center}
\includegraphics[clip,width=3 in, keepaspectratio]{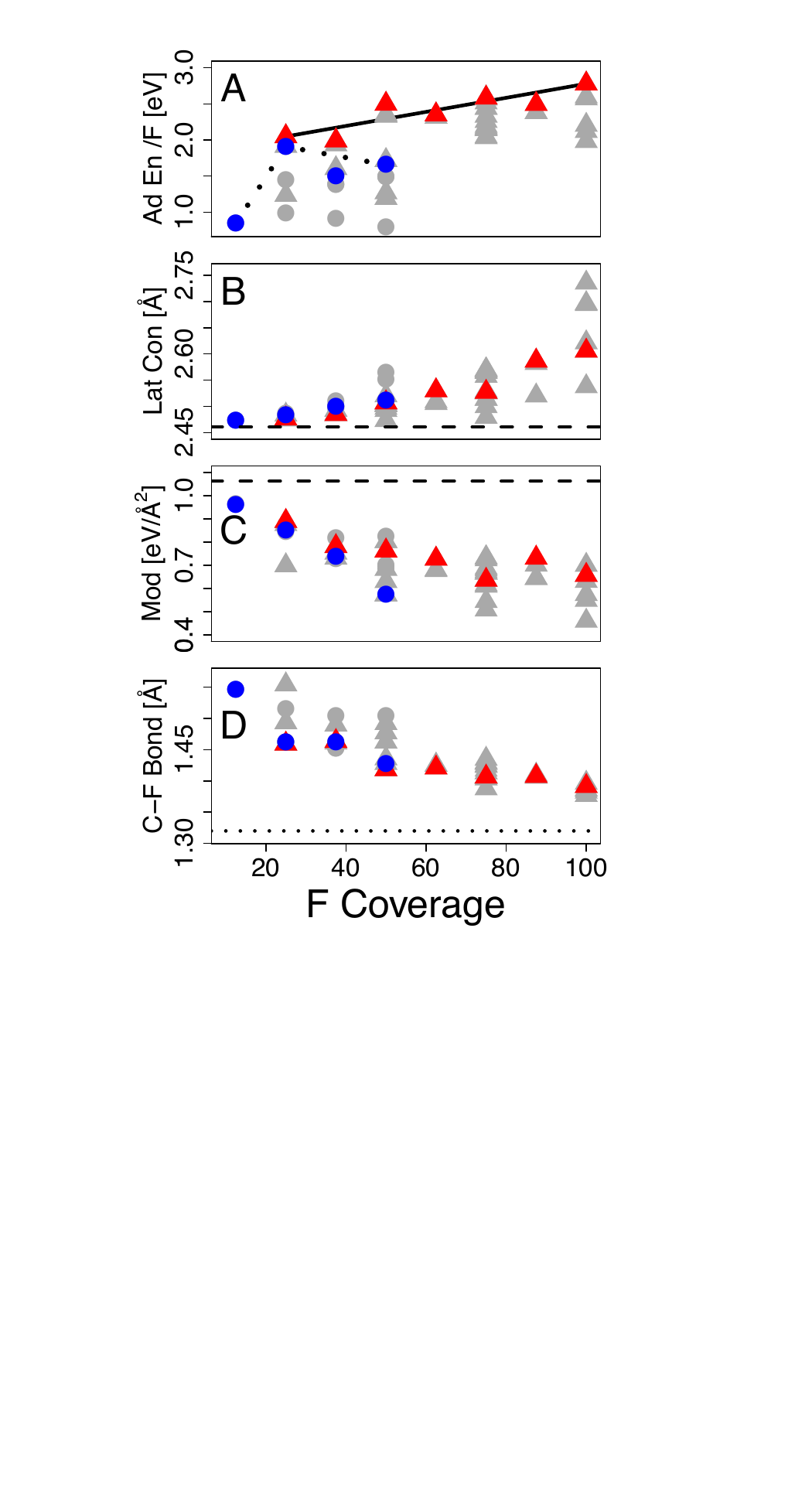}
\end{center}
\caption{\textit{Calculated results as a function of Fluorine coverage.  The gray circles (triangles) represent structures with one-sided (two-sided) fluorination.  The blue circles (red triangles) are the one-sided (two-sided) motifs that have the highest F adsorption energies at a given F coverage.  (A) gives the adsorption energy per F atom, (B) the equilibrium lattice constant, (C) the two-dimensional analog of the bulk modulus, and (D) the C-F bond length.  The dotted (solid) black line in (A) is used to guide the eye in determining the motif of the one-sided (two-sided) structures with the highest adsorption energy per F atom.  The dashed, horizontal lines in (B), (C), and (D) represent the values for pristine graphene.  The dotted line in (D) is the length of a C-F covalent bond.}}
\label{fig:vsF}
\end{figure}

The adsorption energy per F, $E_{A}$, was computed using
\begin{equation}
E_{A} = -(E_{GR\_nF} - (E_{GR} + n_{F} E_{F}))/n_{F}
\label{eq:adsorptionenergy}
\end{equation}
where $E_{GR\_nF}$ is the ground state total energy of the fluorinated graphene, $E_{GR}$ is the ground state total energy of graphene, $E_{F}$ is the energy of a single F atom, and $n_{F}$ is the number of F atoms in the system. From this definition we find that the structures with larger adsorption energies have lower ground state total energies. The adsorption energy per F atom, $E_{A}$ versus coverage is given in Fig.~\ref{fig:vsF} (A) and in Table~\ref{table:configurations} column 6.  We also find that the one-sided coverage with the maximum adsorption energy per F atom is at 25 \% coverage. The dotted line in Fig.~\ref{fig:vsF} (A) shows the maximum adsorption energy for 1-sided motifs.  We also see that the maximum adsorption energies of the two-sided stoichiometries increases (solid line in Fig.~\ref{fig:vsF} (A)) until full coverage (100 \% coverage).  The one-sided F-graphene motif with the highest per F adsorption energy has F atoms at the para position sites (Table~\ref{table:configurations} Item 4)~\cite{Fgrapheneparapostions}, while the fully fluorinated system with the highest adsorption energy is in the motif of Item 51 in Table~\ref{table:configurations}~\cite{Ewels06216103,Sahin11115432}. From modeling of the C K$\alpha$ spectroscopy Bulusheva \textit{et al.} suggested that experimental C$_{2}$F is in a configuration similar to Item 26, which we find to be one of the most energetically favorable C$_{2}$F motifs~\cite{Bulusheva021}.  We note that the adsorption energy per F atom vs coverage is a convenient way of graphically representing other data.  We will use it in Figs.~\ref{fig:conductingstates} and~\ref{fig:polarization} to relate conduction and intrinsic polarization properties to the fluorine coverage and adsorption energies.

We note several points about the calculated adsorption energies.  For stoichiometries that have both one-sided and two-sided motifs, the motif with the highest adsorption energy per F atom is always two-sided.  The one sided (Item 4) and two sided (Item 5) C$_{4}$F coverages with the F in the para positions (adatoms on the C atoms labeled 1 and 6 in Fig.~\ref{fig:configurations}) are similar in energy, but when the F atoms are in the ortho (adatoms at 1 and 2) or meta (adatoms at 1 and 7) positions  the energy difference is larger~\cite{Wade2003}.  This is likely due to the interactions between the F atoms in the ortho and para cases and the relatively smaller interaction in the para configuration.  If we take the motifs with the maximum adsorption energies for each stoichiometry and then break them into two sets, those with an even number of bound fluorine atoms and those with an odd number of bound fluorines, we find that for each set the maximum adsorption energy increases monotonically, whereas taken together, the increase is not monotonic.  Sublattice imbalance in Table 1 is defined as the absolute difference between the number of F atoms on the A sublattice and on the B sublattice.  In cases where a particular stoichiometry allows for different values of sublattice imbalance the motifs that minimize the imbalance are found to have a lower total energy~\cite{Casolo09054704}.

The lattice constant, as found by the Birch-Murnaghan equation of state, of each motif is given in Table~\ref{table:configurations} column 7 and Figure~\ref{fig:vsF}(B).  It shows that the lattice constant tends to increase as the fluorination increases.  This is consistent with charge being drawn out of the C-C bonds by the fluorines.  For pristine graphene we found an equilibrium lattice constant of 2.4607 \AA\, in good agreement with the experimental value of 2.4602 \AA~\cite{Peres10155442}.  For CF, in the motif of Item 51, we found a lattice constant of 2.61 \r{A}, which is near the value calculated by others~\cite{Munoz10368,Zboril102885,Ribas11143,Artyukhov105389}, but it is much larger than the experimental value obtained by Nair \textit{et al.}, 2.48 \r{A}~\cite{Nair102877}.  Artyukhov and Chernozatonskii argue that this difference can be explained by a surface structure like Item 52~\cite{Artyukhov105389}.  Alternatively, it could result from domains of CF of one motif, or different motifs, where the average lattice constant is in the range found experimentally.  Our results also agree well with the C$_{4}$F lattice constant values reported in Ref.~\onlinecite{Ribas11143}.

We performed variable cell optimizations on a representative subset of motifs (the set with the highest adsorption energy per F atom for each stoichiometry) to determine if the fluorination of graphene caused significant changes in the Bravais lattice~\cite{Giannozzi09395502}.  These calculations allow the lattice parameters and the positions of the atoms within the cell to be optimized.  The results gave lattice constants that were approximately equal (to within a few hundredths of an Angstrom) to the values given in Table~\ref{table:configurations} and lattice angles that were consistent with a hexagonal lattice.  Thus, a simple two-dimensional hexagonal lattice is sufficient for describing fluorinated graphene~\cite{Nair102877}.

When we fit the total energy versus crystal volume curve with the Birch-Murnaghan equation of state, one of the parameters that can be found is the bulk modulus.  With fluorinated graphene there is no well-defined way to account for its thickness.  Thus, we cannot accurately compute the bulk modulus.  Therefore, Figure~\ref{fig:vsF} (C) and Table~\ref{table:configurations} gives the two-dimensional version of the bulk modulus as a way to discuss qualitatively the change in the elasticity with fluorination.  These values were computed by fitting changes in the total energy as the surface area changes with the Birch-Murnaghan equation of state.  Pristine graphene has a elastic modulus of 17.2 eV/\AA.  The values computed for fluorinated graphene decrease as fluorination increases.  Sahin \textit{et al.} argue that the rehybridization of the carbon atoms into sp$^{3}$ structures is the main cause of the change in bulk modulus with fluorentation~\cite{Sahin11115432}.

In Table~\ref{table:configurations} column 9 we list the average C-F bond lengths for each motif (shown in Fig.~\ref{fig:vsF} (D)).
Although the theoretical lattice constants in the literature for Item 51 are all similar, there is a wide range of reported theoretical C-F bond lengths~\cite{Ewels06216103,li099274,Sahin11115432,Artyukhov105389} for this structure, with our result near the lower end of the values reported.   Our results are in the middle of the range of reported C-F bond lengths for systems like C$_{2}$F and C$_{4}$F.  Experimental values for these two systems fall over a wide range, from 1.36 \r{A} to 1.64 \r{A}, with our results within the experimental range~\cite{Ebert747841,Kamarchik78296,Ebert747841,Dubois0611800}.  Covalent (ionic) C-F bonds typically have bond lengths of about 1.32 \r{A} (2.01 \r{A}). The values that we obtain for the fluorinated graphene systems, $d_{C-F} = [1.377,1.554]$ indicate that the C-F bond is semi-ionic~\cite{Sahin11115432} at low coverage and becomes more covalent as the fluorination increases.  This agrees with X-ray photoemission spectroscopy, which shows that the bonds are semi-ionic at coverages less then C$_{2}$F, but that they become covalent at higher coverage~\cite{An024235,Wheeler122307}.

The transition from ionic bonding to covalent bonding between carbon atoms in graphene and adsorbates is based on re-hybridization of the carbon atoms.  At lower coverages most of the carbon atoms have sp$^{2}$-like hybridization, but at higher coverages most are in sp$^{3}$-like states.  At low coverages more ionic bonding reduces the strain in the graphene layer that would arise from covalent sp$^{3}$ bonds.  At higher coverages the carbon atoms are coordinated with more than one fluorine atom, which reduces the amount of charge available from the carbon layer to the fluorine and favors covalent bonding.

\begin{figure}
\begin{center}
\includegraphics[clip,width=3 in, keepaspectratio]{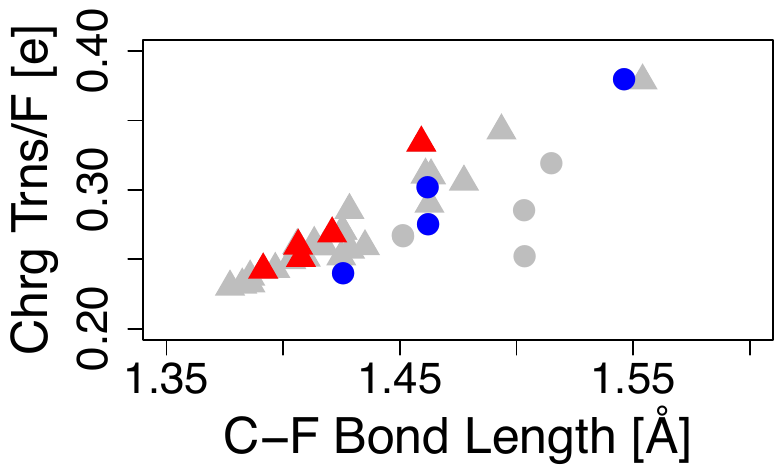}
\end{center}
\caption{Charge transferred into the F atoms versus C-F bond length.  The blue circles (red triangles) represent the one-sided (two-sided) motifs that have the highest adsorption energy per F atom for their respective stoichiometries.}
\label{fig:charge-transfer}
\end{figure}

\begin{figure}
\begin{center}
\includegraphics[clip,width=6 in, keepaspectratio]{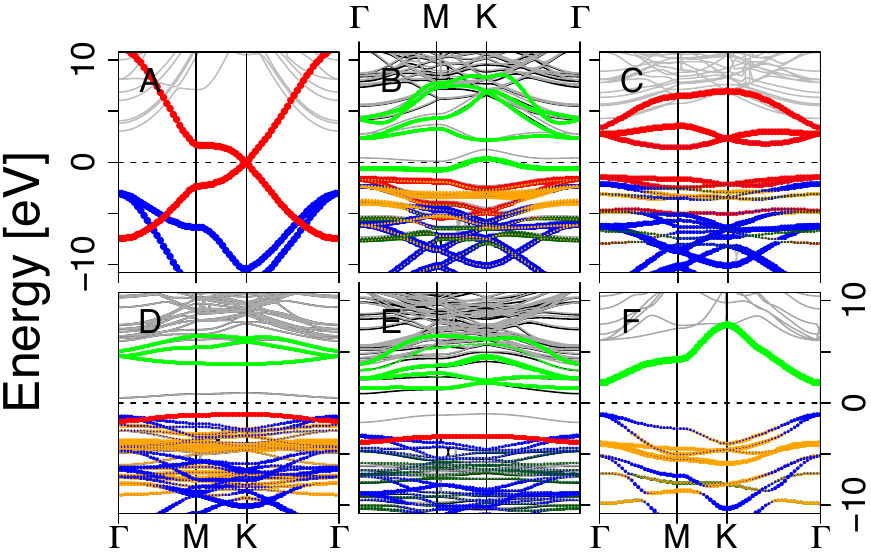}
\end{center}
\caption{\textit{Band dispersions of: [A] graphene, [B] C$_{8}$F (Table~\ref{table:configurations} Item 1), [C] C$_{4}$F (Item 4), [D] C$_{4}$F$_{3}$ (Item 36), [E] C$_{8}$F$_{7}$ (Item 44), [F] CF (Item 52).  The DFT results are given by light gray, dark gray, and black lines; the black lines represent the spin-up results, dark gray the spin-down results, and light gray spin degenerate results.  The maximally localized Wannier function interpolation is illustrated using blue to represent the C$_{sp^{2}}$ bonds, red the non-bonded C$_{p_{z}}$, dark green the symmetric bonds between C$_{p_{z}}$ and F$_{p_{z}}$, green the antisymmetric bonds between C$_{p_{z}}$ and F$_{p_{z}}$, and orange the F$_{p_{xy}}$ bonds.  For clarity, [A] and [F] are for 1-by-1 supercells, giving fewer bands than in the other figures.}}
\label{fig:bandstructures}
\end{figure}
\begin{figure}
\begin{center}
\includegraphics[clip,width=3 in, keepaspectratio]{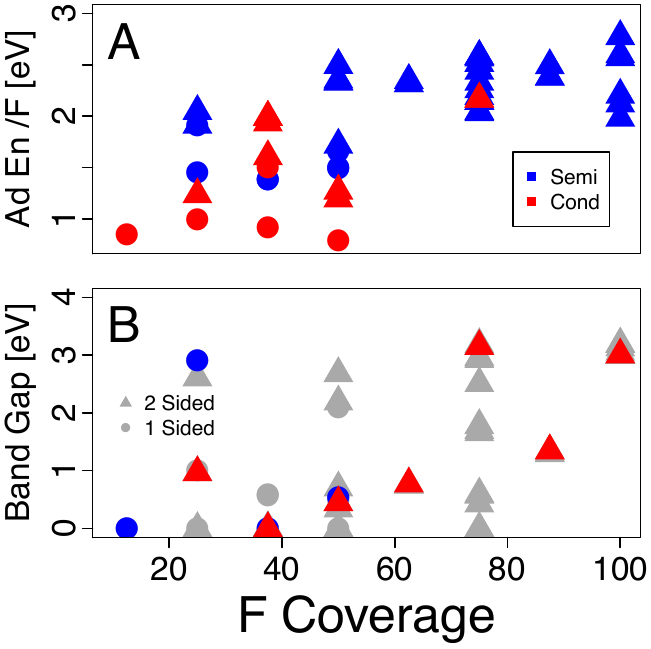}
\end{center}
\caption{(A) Plot of semiconducting (blue) and metallic (red) systems. (B) Distribution of band gaps versus coverage.  The blue circles (red triangles) are the one-sided (two-sided) motifs that have the highest adsorption energy per F atom.}
\label{fig:conductingstates}
\end{figure}
\begin{figure}
\begin{center}
\includegraphics[clip,width=2 in, keepaspectratio]{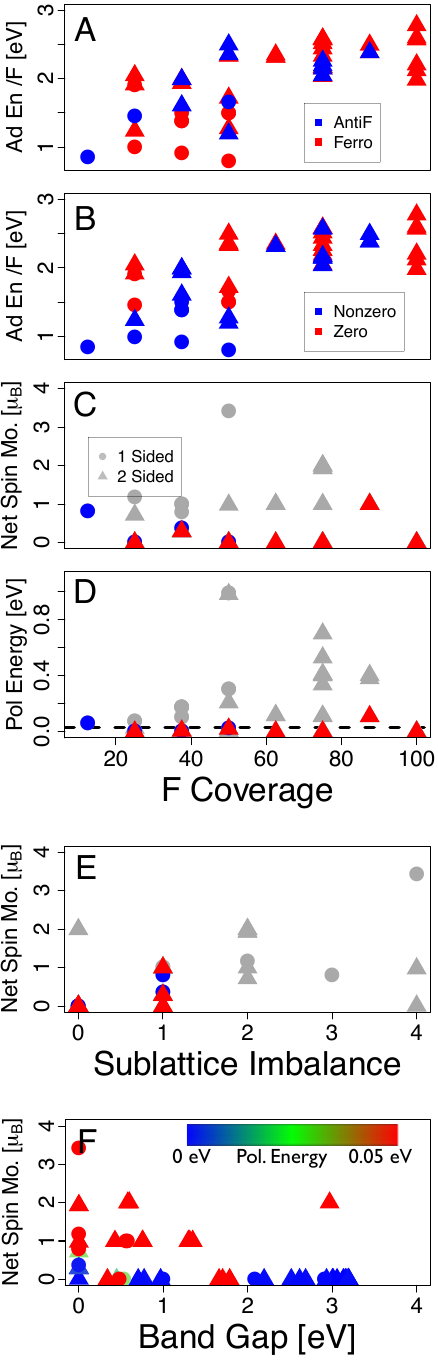}
\end{center}
\caption{Magnetic Properties: (A) Motifs with ferromagnetic (red) and antiferromagnetic (blue) spin alignments in the unit cell.  (B) Motifs with net spin moment zero (red) and non-zero (blue). (C-E) The gray circles (triangles) represent structures with one-sided (two-sided) fluorination.  The blue circles (red triangles) are the one-sided (two-sided) motifs that have the highest F adsorption energies.  (C) Net spin moment versus fluorine coverage. (D) Spin moment energy versus fluorine coverage. (E) Spin-moment versus the difference in the number of F adatoms on each sublattice (sublattice imbalance). (F) Net spin moment versus band gap where the color of the symbols represents the polarization energy.}
\label{fig:polarization}
\end{figure}

Projecting the DFT results for electronic properties onto tight-binding bands can give physical insight into binding and electrical properties.  We do this projection onto a set of maximally localized Wannier functions (MLWF).  Figure~\ref{fig:bandstructures} gives the DFT band dispersion plots along with the corresponding MLWF orbital projections of pristine graphene and of several illustrative fluorinated graphene systems.
In the Wannierization of the DFT results the high symmetry energy bands are disentangled separately in the following energy windows: the valence band, the localized states in the band gap, and the lowest energies of the conduction band.  In the valence band, there are branches coming from s-like states centered on F atoms that are separated at very low energies ($< -10$~eV). The other bands are disentangled in terms of the following orbitals (Wannier states): one s-like orbital centered on each C-C pair ($\sigma$ bond) including both $sp^{2}$ (non-fluorinated) and $sp^{3}$ (fluorinated) carbons, which produce deep energy states; one $s$-like orbital centered on each C-F pair, also with a low energy in the valence band; one $p_z$-like orbital for each pair of non-fluorinated neighboring C atoms ($\pi$ bond) at the top of the valence band; two $p_x$, $p_y$ orbitals on each F atom accounting for the paired electrons in the $p$-shells of the F atoms and giving localized (non-dispersive), low energy states.  For the conduction bands we interpolate only the branches in a small low-energy window that contains the bands that are most relevant for transport. Higher states cannot be disentangled with tight-binding orbitals because they hybridize at high energies with an unlimited number of quasi-free surface states. We find that the lowest conduction branches are given by $p_z$-like orbitals centered on C-F pairs (antibonding $\sigma^{*}$ states) and by a combination of $d_{xz}$ and $d_{yz}$-like orbitals centered on unfluorinated C-C pairs (antibonding $\pi^*$ states). For fully fluorinated graphene the conduction band is made up solely by $p_z$-like orbitals centered on C-F pairs.

Figures~\ref{fig:bandstructures} (A) shows the band dispersion of pristine graphene.  The HOMO and LUMO states of graphene are composed entirely of the $\pi$ and $\pi^{*}$ molecular orbitals made from C$p_{z}$ orbitals.  Lower energy valence branches are composed entirely of $sp^2$ molecular orbitals.  Due to the symmetry of pristine graphene, there is no hybridization between its $\pi$ and $sp^2$ orbitals; therefore its energy bands are not entangled and can be interpolated in a single energy window that includes the lowest conduction bands.  There are also three high-energy conduction branches created by anti-bonding C $sp^2$ orbitals and a set of quasi-parabolic branches from states outside the graphitic plane.

Figure~\ref{fig:bandstructures}B-F deal with systems having multiple F adatoms.  Bonds formed between the F $p_{z}$ and the C $p_{z}$ orbitals force the C atoms into $sp^3$ hybridization, breaking the symmetry of graphene. In many cases this opens a gap via the formation of new C-F bonding states deep in the valence band and C-F antibonding states in the conduction band~\cite{Sahin11115432}, as can be seen in Figs.~\ref{fig:bandstructures}. The resulting bands are entangled due to orbital hybridization, manifested as anticrossings between branches in the Brillouin zone. As the degree of fluorination is increased, the number of C-C $\pi$ states at the top of the valence band is decreased and the number of deep C-F $\sigma$ bonds is increased, until all $\pi$ bonds are removed for CF and the top of the valence band is given by $\sigma$ bonds.

The C$_{4}$F structure (Item 4) in Fig.~\ref{fig:bandstructures} (C) is an exception to this trend.  The HOMO and LUMO states can be described completely by the hybridization of the C$_p{_{z}}$ of the carbon atoms that are not bound to F atoms and the $sp^{2}$ bonding between carbon atoms, although the latter orbitals are only weakly hybridized. The difference in electronic structures between the C$_{4}$F structure discussed here and the other fluorinated graphene structures in this paper occurs because the arrangement of F atoms allows for both an aromatic ring and equal numbers of adsorbates on the graphitic sublattices.

For an odd number of non-fluorinated carbons in the supercell the result is a spin-polarized band structure.  These bands are interpolated by $p_{z}$-like Wannier states centered on the unfluorinated C atoms. For example, the C$_8$F phase has an unpaired $p_{z}$ orbital surrounded by three pairs of $\pi$ bonds that give the valence band in Fig.~\ref{fig:bandstructures} (B), and the C$_8$F$_7$ has an unpaired $p_z$ orbital surrounded by $sp^3$ carbons as in Fig.~\ref{fig:bandstructures} (E).

Table~\ref{table:configurations} column 11 lists the band gaps of the systems.  The letter C is listed instead of a numerical value in cases where the system's Fermi level is within a band.  Figure~\ref{fig:bandstructures} (B) shows such a case.  Many of the systems have indirect band gaps, and thus will be less optically active than the systems with direct gaps.

We find that the majority of the configurations studied here are semiconductors.  Figure~\ref{fig:conductingstates} (A) shows how the metallic and semiconducting configurations are distributed as functions of the adsorption energy and coverage.  The semiconducting systems tend to have larger adsorption energies per F, suggesting that it may be hard to produce a highly fluorinated system that is metallic.  We found that a filled valence band occurred only when the carbon atoms without an attached F were in pairs or rings.  Figure~\ref{fig:conductingstates} (B) shows the distribution of band gaps versus coverage.  We see that it is possible to tune the band gap over a wide range by F adsorption.

Trends in the local spin polarizations of fluorinated graphene can be understood in terms of the bipartite lattice of graphene and Hund's rule.  Each carbon atom in graphene forms three covalent bonds ($\sigma$ molecular orbitals) leaving one electron to be part of the $\pi$ bonding.  If one C atom has a spin up electron in it's $p_{z}$-orbital then it's neighbors will have spin down electrons in their p$_{z}$-orbitals.  The formation of the $\pi$ conjugation effectively makes the spin polarization of the graphene sheet zero.
Each neutrally charged F atom will have a spin $\pm 1/2$ excess before adsorption on to the surface.  To satisfy Hund's rule a fluorine's $p_{z}$ electron pairs up with a spin state opposite to that of the C atom on which it adsorbs.
As the F adsorbs onto a C atom it forms a semi-ionic bond, the covalent portion of which kills the majority of the spin polarization in the F atom while only slightly changing the spin polarization of the C atom.  The formation of the C-F bond breaks the $\pi$-conjugate bonding of the C atom with its neighbors.  The nearest neighbors, which are the most disturbed by the disrupted $\pi$ conjugation, share the majority of the spin excess~\cite{Suarez2012,Matte099982}.  The bipartite nature of the surface ensures that the induced spin polarization of the nearest neighbors is shared with the other C atoms that make up the same sublattice~\cite{Casolo09054704}.  Further F adsorptions, depending on placement, produce strong or weaker induced polarizations in individual atoms.

Spin-unpolarized and spin-polarized calculations were made. The spin-polarized case consisted of two calculations for each motif, one with an antiferromagnetic initial spin alignment in the 2-by-2 unit cell and the other with ferromagnetic alignment, as described in Section II.   Total energies then determined the ground state.  Figure~\ref{fig:polarization} (A) shows the distribution of ferromagnetic and antiferromagnetic spin alignments among the ground state configurations of the unit cells, and (B) shows the distribution of motifs that have zero and non-zero net spin moments.  Analysis of the ground states of the different motifs gives a range of spin alignments, including nonmagnetic, ferrimagnetic, and antiferromagnetic.  Ferrimagnetic spin alignments, where the carbon atoms without adsorbates may all have different spin moments, are the most common.  Items 17 and 20 are an interesting case study.  Both have four F atoms bonded to C atoms 1, 3, 5, and 7.   All of the F atoms are on the same side of the graphene sheet in Item 17, while Item 20 has F atoms on both sides.  In both Items 17 and 20 the ground state has ferromagnetic alignment.  Item 17 has a large net magnetic moment, and Item 20 has an almost zero net moment.  Thus the net magnetic moment of the unit cell is sensitively dependent on the details of the fluorinated system, such as the structure, coverage, and arrangement of the adsorbates.~\cite{rivera2014ab}

Figures~\ref{fig:polarization} (C) and (E) show the net spin moment versus coverage and versus the sublattice imbalance, respectively.  Zhou \textit{et al.} calculated the properties of hydrogenated graphene with stoichiometries of C$_4$H, C$_2$H, and CH, and they speculate that, ``only half hydrogenation can introduce magnetization"~\cite{Zhou093867}. This is not the case for fluorinated graphene. We find that there is at least one motif for every stoichiometry, other than fully fluorinated graphene, that has a nonzero intrinsic spin moment.  Also, we find that very few of the structures have a net spin moment that is approximately an integer multiple of one Bohr magnetron.


We also estimated the polarization energy, $E_{P}$, of the fluorinated graphene systems (Figure~\ref{fig:polarization} (D) and Table~\ref{table:configurations} column 16).  $E_{p}$ is given by $E_{p} = E_{sr} - E_{su}$, where $E_{sr}$ is the energy of the system when computed in a spin restricted calculation, and $E_{su}$ is the energy of the system when in the ferromagnetic alignment (or antiferromagnetic alignment, whichever is energetically favorable).  As seen in the last column of Table~\ref{table:configurations}, many of the systems have $E_{P}$ values that are greater than 0.03 eV making it possible for them to have stable spin moments at room temperature.

Figure~\ref{fig:polarization} (F) gives magnetic moments of the unit cells versus band gaps of the systems, where the colors of the  points provide estimate of polarization energy (all points with a polarization energy greater than or equal to 0.05 eV are red).
The only motif with a net spin moment above 2 $\mu_{B}$, Item 17, has a band structure like a highly p-doped semiconductor and has a stable spin moment at room temperature~\cite{Zhang12201409}.  There are five motifs that have a spin moment near 2 $\mu_{B}$, with band structures ranging from a 3 eV band gap to a system is metallic Fig.~\ref{fig:polarization} (F)).  All five have C$_{4}$F$_{3}$ stoichiometry, and all have a spin moment that is thermally stable above room temperature.  The range of band gaps is smaller for motifs with spin polarizations near 1 $\mu_{B}$.  Thus, it may be possible to produce motifs that have desirable electrical and spin polarization properties that are thermally stable.


\section{Summary}\label{sec:Summary}

We have studied graphene systems for a wide range of fluorinations within density functional theory.  Several atomic arrangements are possible for most stoichiometries.  Wide ranges of structural, electronic, elastic, and magnetic properties have been calculated for them.  Systems vary from metallic to semiconducting with widely varying band gaps.  Electronic properties and lattice constants vary with coverage, and a rich range of antiferromagnetic and ferromagnetic polarizations is found.  It is reasonable to expect that a variety of these phases may occur in real materials systems.

\section{Acknowledgment}\label{sec:Acknowledgment}

This work was supported by the Office of Naval Research.  Computer resources were provided by the DoD High Performance Computing Modernization Program.  C.E.J. acknowledges support of the NRL/NRC  National Research Council Postdoctoral Associateship.  We thank J. Robinson, S. Schmucker, and P. Dev for valuable discussions concerning this work.

%

\bibliographystyle{unsrt}

\begin{thebibliography}{10}
\bibitem{labelingexample}
A quick example will explain this. Items 4 and 7 both have two F adatoms in the
supercell. Item 4 has F atoms bonded to C atoms 1 and 6 with both on the same
side of the graphene sheet. In Item 7 the F are bonded to the same atoms, but
one F is above the graphene sheet (sticking out of the page in Fig. 1), and
the other is below the graphene sheet (pointing into the page).
\bibitem{spinordering}
In a few cases (e.g., Item 16) all of the F atoms are attached to the same
sublattice, and thus our scheme could not produce an antiferromagnetic
ordering. In these cases we alternated the initial spin surplus on the F
atoms. In one case, Item 1, we alternated the initial spin surplus on the C
atoms to produce an antiferromagnetic structure.
\bibitem{Fgrapheneparapostions}
As a check on this statement we also computed the absorption energies of two F
atoms in meta, ortho, and para configurations in the middle of a much larger
supercell that was fully relaxed, but otherwise free of defects. The systems
consisted of rectangular graphene supercells that were 29.58 \AA\, along the
x-direction and 25.6176 \AA\, along the y-direction, where the x-direction
was along a zig-zag pattern of the graphene sheet. The absorption energy per
F atom of the meta, ortho, and para conformations were 2.679 eV, 2.420 eV,
and 2.826 eV, respectively.
\end{thebibliography}



\begin{turnpage}
\begin{ruledtabular}
\begin{table}
\centering
{\scriptsize \begin{tabular}{|c|c|c|c|c|c|c|c|c|c|c|c|c|c|c|}
\hline
Item  & Stoichiometry & Coverage & Above & Below & $E_{A}$ [eV] & a [\AA] &  BM & $d_{C-F}$ [\AA]  & Charge [e/F] & Gap [eV] &  F/A  & Imbalance & Net Moment [Bohr mag.]& Pol. En. [eV]\\
\hline
1	&	C$_{8}$F	&	12.5	&	2	&	NA	&	0.845	&	2.473	&	0.962	&	1.546	&	0.380	&	C	&	A	&	1	&	-0.81	&	0.061	\\
2	&	C$_{4}$F	&	25.0	&	1,2	&	NA	&	1.446	&	2.484	&	0.845	&	1.461	&	0.601	&	0.997	&	A	&	0	&	0	&	0.001	\\
3	&	C$_{4}$F	&	25.0	&	1,3	&	NA	&	0.990	&	2.486	&	0.875	&	1.515	&	0.319	&	C	&	F	&	2	&	1.17	&	0.071	\\
4	&	C$_{4}$F	&	25.0	&	1,6	&	NA	&	1.908	&	2.483	&	0.849	&	1.462	&	0.302	&	2.909	&	F	&	0	&	0	&	0.001	\\
5	&	C$_{4}$F	&	25.0	&	1	&	6	&	1.915	&	2.484	&	0.699	&	1.494	&	0.343	&	2.613	&	F	&	0	&	0	&	0.001	\\
6	&	C$_{4}$F	&	25.0	&	1	&	3	&	1.237	&	2.474	&	0.876	&	1.554	&	0.379	&	C	&	F	&	2	&	0.72	&	0.028	\\
7	&	C$_{4}$F	&	25.0	&	1	&	2	&	2.048	&	2.476	&	0.889	&	1.459	&	0.334	&	0.968	&	F	&	0	&	0	&	0.001	\\
8	&	C$_{8}$F$_{3}$	&	37.5	&	1,2,5	&	NA	&	1.499	&	2.500	&	0.738	&	1.462	&	0.275	&	C	&	F	&	1	&	0.36	&	0.003	\\
9	&	C$_{8}$F$_{3}$	&	37.5	&	1,3,5	&	NA	&	0.908	&	2.510	&	0.817	&	1.503	&	0.285	&	C	&	F	&	3	&	0.79	&	0.101	\\
10	&	C$_{8}$F$_{3}$	&	37.5	&	2,3,6	&	NA	&	1.378	&	2.499	&	0.728	&	1.451	&	0.267	&	0.576	&	F	&	1	&	1	&	0.173	\\
11	&	C$_{8}$F$_{3}$	&	37.5	&	1,2,3	&	NA	&	1.378	&	2.496	&	0.730	&	1.451	&	0.267	&	0.559	&	F	&	1	&	1	&	0.172	\\
12	&	C$_{8}$F$_{3}$	&	37.5	&	1,2	&	6	&	1.607	&	2.493	&	0.732	&	1.490	&	0.950	&	C	&	A	&	1	&	-0.28	&	0.006	\\
13	&	C$_{8}$F$_{3}$	&	37.5	&	1,3	&	6	&	1.936	&	2.486	&	0.753	&	1.461	&	0.311	&	C	&	F	&	1	&	0.29	&	0.009	\\
14	&	C$_{8}$F$_{3}$	&	37.5	&	1,6	&	3	&	1.991	&	2.485	&	0.781	&	1.463	&	0.310	&	C	&	F	&	1	&	0.29	&	0.007	\\
15	&	C$_{8}$F$_{3}$	&	37.5	&	1,6	&	2	&	1.991	&	2.485	&	0.782	&	1.463	&	0.932	&	C	&	A	&	1	&	-0.29	&	0.007	\\
16	&	C$_{2}$F	&	50.0	&	2,3,6,7	&	NA	&	1.656	&	2.512	&	0.694	&	1.427	&	0.978	&	0.530	&	A	&	0	&	0	&	0.028	\\
17	&	C$_{2}$F	&	50.0	&	1,3,5,7	&	NA	&	0.788	&	2.550	&	0.823	&	1.504	&	0.252	&	C	&	F	&	4	&	3.42	&	0.302	\\
18	&	C$_{2}$F	&	50.0	&	1,2,5,6	&	NA	&	1.493	&	2.564	&	0.683	&	1.426	&	0.240	&	2.083	&	F	&	0	&	0	&	0.001	\\
19	&	C$_{2}$F	&	50.0	&	1,3,5	&	7	&	1.196	&	2.497	&	0.683	&	1.492	&	1.182	&	C	&	A	&	4	&	0.98	&	0.209	\\
20	&	C$_{2}$F	&	50.0	&	1,7	&	3,5	&	1.272	&	2.473	&	0.630	&	1.477	&	0.306	&	C	&	A	&	4	&	-0.01	&	0.001	\\
21	&	C$_{2}$F	&	50.0	&	3,7	&	2,6	&	2.498	&	2.507	&	0.803	&	1.418	&	1.130	&	0.447	&	A	&	0	&	0	&	0.021	\\
22	&	C$_{2}$F	&	50.0	&	2,5	&	1,6	&	2.355	&	2.502	&	0.763	&	1.436	&	1.144	&	2.684	&	A	&	0	&	0	&	0.001	\\
23	&	C$_{2}$F	&	50.0	&	2,6	&	1,5	&	2.338	&	2.493	&	0.573	&	1.428	&	0.285	&	2.192	&	F	&	0	&	0	&	0.001	\\
24	&	C$_{2}$F	&	50.0	&	5,6	&	1,2	&	1.719	&	2.521	&	0.582	&	1.463	&	0.290	&	0.702	&	F	&	0	&	0	&	0.001	\\
25	&	C$_{2}$F	&	50.0	&	1,2,3,4	&	NA	&	1.656	&	2.508	&	0.702	&	1.427	&	0.979	&	0.477	&	F	&	0	&	0	&	0.987	\\
26	&	C$_{2}$F	&	50.0	&	1,3	&	2,4	&	2.497	&	2.503	&	0.803	&	1.418	&	1.130	&	0.338	&	F	&	0	&	0	&	0.984	\\
27	&	C$_{8}$F$_{5}$	&	62.5	&	5,6	&	4,7,8	&	2.319	&	2.510	&	0.678	&	1.425	&	0.270	&	0.756	&	F	&	1	&	1	&	0.118	\\
28	&	C$_{8}$F$_{5}$	&	62.5	&	5,7	&	4,6,8	&	2.351	&	2.530	&	0.727	&	1.421	&	0.268	&	0.772	&	F	&	1	&	0	&	0.001	\\
29	&	C$_{8}$F$_{5}$	&	62.5	&	5,6	&	3,7,8	&	2.318	&	2.506	&	0.681	&	1.425	&	0.270	&	0.754	&	F	&	1	&	1	&	0.118	\\
30	&	C$_{4}$F$_{3}$	&	75.0	&	2,3,6,7	&	5,8	&	2.441	&	2.514	&	0.611	&	1.413	&	0.262	&	3.188	&	F	&	0	&	0	&	0.001	\\
31	&	C$_{4}$F$_{3}$	&	75.0	&	1,5,6	&	4,7,8	&	2.583	&	2.527	&	0.617	&	1.406	&	0.260	&	3.162	&	F	&	0	&	0	&	0.001	\\
32	&	C$_{4}$F$_{3}$	&	75.0	&	2,5,6	&	4,7,8	&	2.168	&	2.529	&	0.667	&	1.435	&	0.259	&	C	&	F	&	2	&	1.93	&	0.114	\\
33	&	C$_{4}$F$_{3}$	&	75.0	&	3,5,7	&	4,6,8	&	2.570	&	2.558	&	0.642	&	1.406	&	0.258	&	2.968	&	A	&	0	&	-2	&	0.404	\\
34	&	C$_{4}$F$_{3}$	&	75.0	&	2,3,7	&	4,5,8	&	2.261	&	2.526	&	0.508	&	1.388	&	1.592	&	1.701	&	A	&	0	&	0	&	0.338	\\
35	&	C$_{4}$F$_{3}$	&	75.0	&	2,6,7	&	4,5,8	&	2.168	&	2.480	&	0.545	&	1.418	&	0.259	&	C	&	F	&	2	&	1.93	&	0.111	\\
36	&	C$_{4}$F$_{3}$	&	75.0	&	3,6,7	&	4,5,8	&	2.517	&	2.500	&	0.637	&	1.406	&	0.260	&	2.514	&	F	&	0	&	0	&	0.001	\\
37	&	C$_{4}$F$_{3}$	&	75.0	&	2,5,6	&	3,7,8	&	2.583	&	2.527	&	0.636	&	1.407	&	0.260	&	3.154	&	F	&	0	&	0	&	0.001	\\
38	&	C$_{4}$F$_{3}$	&	75.0	&	2,5	&	3,4,7,8	&	2.049	&	2.570	&	0.688	&	1.424	&	1.523	&	1.785	&	A	&	0	&	0	&	0.531	\\
39	&	C$_{4}$F$_{3}$	&	75.0	&	2,6	&	3,4,7,8	&	2.036	&	2.565	&	0.674	&	1.425	&	0.252	&	0.580	&	F	&	2	&	2	&	0.410	\\
40	&	C$_{4}$F$_{3}$	&	75.0	&	5,6	&	3,4,7,8	&	2.334	&	2.567	&	0.667	&	1.404	&	0.249	&	2.931	&	F	&	0	&	0	&	0.001	\\
41	&	C$_{4}$F$_{3}$	&	75.0	&	3,5,7	&	2,4,8	&	2.194	&	2.567	&	0.724	&	1.430	&	1.586	&	1.660	&	A	&	0	&	0	&	0.702	\\
42	&	C$_{4}$F$_{3}$	&	75.0	&	5,7	&	2,4,6,8	&	2.147	&	2.570	&	0.733	&	1.429	&	1.541	&	0.427	&	F	&	2	&	1	&	0.412	\\
43	&	C$_{4}$F$_{3}$	&	75.0	&	3,7	&	2,4,6,8	&	2.147	&	2.570	&	0.732	&	1.429	&	0.257	&	0.597	&	F	&	2	&	2	&	0.403	\\
44	&	C$_{8}$F$_{7}$	&	87.5	&	1,3,5,7	&	4,6,8	&	2.495	&	2.587	&	0.701	&	1.408	&	0.251	&	1.347	&	F	&	1	&	1	&	0.110	\\
45	&	C$_{8}$F$_{7}$	&	87.5	&	2,5,6	&	3,4,7,8	&	2.384	&	2.582	&	0.644	&	1.406	&	1.727	&	1.312	&	A	&	1	&	-1	&	0.382	\\
46	&	C$_{8}$F$_{7}$	&	87.5	&	2,3,6,7	&	1,4,5	&	2.384	&	2.520	&	0.628	&	1.410	&	0.251	&	1.297	&	F	&	1	&	1	&	0.405	\\
47	&	CF	&	100.0	&	1,2,3,5,6	&	4,7,8	&	2.206	&	2.695	&	0.550	&	1.382	&	0.232	&	3.193	&	F	&	0	&	0	&	0	\\
48	&	CF	&	100.0	&	1,2,4,5,8	&	3,6,7	&	1.982	&	2.696	&	0.460	&	1.386	&	0.232	&	3.131	&	F	&	0	&	0	&	0	\\
49	&	CF	&	100.0	&	1,2,5,6	&	3,4,7,8	&	2.573	&	2.622	&	0.701	&	1.386	&	0.237	&	3.010	&	F	&	0	&	0	&	0	\\
50	&	CF	&	100.0	&	1,3,5,6,7	&	2,4,8	&	2.121	&	2.736	&	0.575	&	1.377	&	0.230	&	3.194	&	F	&	0	&	0	&	0	\\
51	&	CF	&	100.0	&	1,3,5,7	&	2,4,6,8	&	2.778	&	2.606	&	0.657	&	1.391	&	0.242	&	3.056	&	F	&	0	&	0	&	0	\\
52	&	CF	&	100.0	&	2,3,6,7	&	1,4,5,8	&	2.601	&	2.538	&	0.631	&	1.397	&	0.243	&	3.001	&	F	&	0	&	0	&	0	\\
\hline
\end{tabular}}
\caption{\textit{Listing of the motifs and physical properties for all of the fluorinated graphene structures studied.  Throughout the paper we will reference a particular structure by its Item number given in the first column.  In order from left to right the columns contain: the item number, stoichiometry, percentage of C atoms with F adatoms, listing of C atoms with F atom above them, listing of C atoms with F atom attached below the graphene sheet, adsorption energy, lattice constant, the one dimensional analog of the bulk modulus, C-F bond length, the charge transfer per F atom, band gap, ground state polarization, sublattice imbalance, net polarization, and the polarization energy.}}
\label{table:configurations}
\end{table}
\end{ruledtabular}
\end{turnpage}

\end{document}